# The evolution of extreme spatial heterochiasmy in amphibians: interactions between sex-specific meiotic program and chromosome architecture

Ezgi Unal[1†], Paris Veltsos[1], Wen-Juan Ma[1†]

[1.] Research group of Ecology, Evolution and Genetics, Biology Department, Vrije Universiteit Brussel, Brussels, Belgium

[†]Correspondence:

Wen-Juan.Ma@vub.be

Hatice.Ezgi.Unal@vub.be

## Abstract

Heterochiasmy, sex differences in meiotic recombination, is usually described in terms of genetic-map length, potentially obscuring more consequential differences in crossover location. Amphibians provide striking examples of extreme spatial heterochiasmy, with male recombination concentrated near chromosome ends and female recombination broadly distributed, yet the evolution and genetic basis of this pattern remains poorly understood. Here, we combine a review with a phylogenetic comparative synthesis of male chiasma location across 97 amphibian species. Male crossover localization is phylogenetically structured, and ancestral reconstruction favors predominantly interstitial recombination at deeper anuran nodes, although not conclusively. The strongest correlate of male terminal crossover location is chromosome morphology: telocentric-rich karyotypes retain more interstitial male chiasmata, whereas metacentric chromosomes show more terminal male chiasmata. Genome size and sex-chromosome system show little association. We propose a sequential model in which sex-specific meiotic regulation first establishes a crossover-precursor placement bias, while centromere position influences which precursors mature into crossovers. This model predicts that chromosome fusion, fission or centromere repositioning can alter male chiasma location without requiring changes to the underlying meiotic program. Our synthesis highlights crossover location as a distinct dimension of heterochiasmy with consequences for genome-wide linkage, adaptation, divergence and sex-chromosome evolution.

## Keywords

Spatial heterochiasmy, recombination landscape, sex-specific meiotic program, chromosome morphology, chiasma location, sex chromosome evolution, amphibians

## 1. Heterochiasmy as variation in recombination landscape

Meiotic recombination generates new combinations of alleles while also performing the essential cytological role of homologous chromosome segregation. A subset of meiotic recombination events is resolved as crossovers (COs), which reshuffle allelic combinations among linked loci and thereby influence the efficacy of natural selection [1,2]. COs are also important for accurate chromosome segregation during meiosis I by providing physical connections between homologous chromosomes [3]. Despite these conserved functions, both CO number and chromosomal distribution vary extensively among species, populations, individuals, sexes, and genomic regions [4,5].

One major source of this variation is heterochiasmy, the differences in recombination between males and females, which has at least two dimensions. The sexes can differ in the total amount of recombination, commonly measured by genetic map length (or CO number), and in their spatial distribution along chromosomes (or recombination topology). These dimensions need not covary: similar genome-wide map lengths can conceal profoundly different chromosomal distributions of COs, whereas large differences in map length do not necessarily imply equally large differences in CO localization. Heterochiasmy therefore cannot be adequately represented by a single female to male map-length ratio [6,7]. A comparative survey of 51 eukaryotes with sex-specific recombination landscapes reinforces this distinction: male recombination is frequently enriched toward chromosome ends whereas female recombination is more broadly distributed, particularly across vertebrates, although many exceptions occur [7]. Thus, sex differences in total recombination and in chromosomal localization are evolutionarily related but not interchangeable traits.

At one extreme of sex-specific recombination variation lies achiasmy, in which meiotic crossover is absent in one sex. Sex-specific achiasmy is strongly associated with the heterogametic sex, a pattern known as the Haldane–Huxley rule. However, this association breaks down when both sexes recombine but differ in their overall recombination rates (quantitative heterochiasmy) [8,9]. Achiasmy should therefore be viewed as a distinct boundary condition rather than the inevitable endpoint of progressively stronger heterochiasmy. Between modest heterochiasmy and complete achiasmy are organisms in which both sexes retain COs but recombination in one sex is strongly restricted to chromosomal regions. We refer to this as extreme spatial heterochiasmy, and it has been observed in both animals and plants. In threespine sticklebacks, female maps are only 1.64-fold longer than male maps, yet male COs are strongly concentrated near chromosome ends and depleted around centromeres, whereas female COs are considerably more evenly distributed [10]. In the dioecious plant *Rumex hastatulus*, female maps are on average only approximately 1.4-fold longer, but males contain continuous low-recombination regions spanning much of the genome [11]. Thus, the chromosomal regions that are effectively reshuffled can differ profoundly between the sexes even when their total genetic map lengths are relatively similar.

Amphibians provide an especially informative system in which to investigate the spatial component of heterochiasmy, particularly sex differences in CO localization along chromosomes. Cytological studies spanning several decades consistently contrast broadly distributed female chiasmata and predominantly terminal male chiasmata, the latter producing characteristic ring-shaped bivalents [12,13]. Modern linkage mapping has confirmed that male recombination is strongly restricted to chromosome ends in distantly related anuran lineages and has revealed that, when comparing species, the magnitude of sex differences in total map length is considerably greater than the spatial restriction of male recombination. For example, among XY species, female autosomal maps in *Hyla arborea* were more than 14-fold longer than male maps [14]. In *Rana temporaria*, the female map was only approximately 3.4-fold longer, yet male COs were almost entirely restricted to chromosome ends, while female COs were broadly distributed along chromosomes [15]. Similarly, *Litoria serrata* shows only a 1.29-fold female to male map-length difference, despite extensive male-specific recombination cold spots across chromosome interiors [16].

Extreme heterochiasmy is not restricted to XY systems. In the ZW tree frogs *Hyla sarda* and *H. savignyi*, female maps are approximately twice as long as male maps, yet males retain strongly terminally restricted recombination despite females being the heterogametic sex [17]. Terminally restricted male recombination is therefore not simply a consequence of male heterogamety. Similar decoupling between total map-length differences and crossover localization is seen in other amphibians: female to male recombination ratios in *Xenopus* range from 1.32 in *X. tropicalis* to 2.66 in *X. borealis* [18]. Together, these studies indicate that total map-length differences between sexes alone do not capture the extent of spatial heterochiasmy, and that sex-specific crossover localization can persist independently of the heterogametic sex.

As females and males share the same centromere position, chromosome-arm configuration and dimensions for a given autosome, chromosome structure alone cannot generate heterochiasmy. Instead, these features may strengthen, weaken or redistribute an underlying sex-specific crossover bias. We propose that extreme spatial heterochiasmy arises through the interaction between a sex-specific meiotic program that biases crossover placement and chromosome-specific structural features that modify how this bias is expressed. Evolutionary changes in centromere position, arm configuration or chromosome dimensions could therefore alter the spatial extent of male recombination without requiring the underlying meiotic program to evolve anew, providing a mechanism by which chromosome evolution can shape the expression and evolutionary dynamics of spatial heterochiasmy.

Amphibians are particularly suitable for testing this hypothesis because they show extensive variation in chromosome morphology and karyotype architecture, especially centromere position and chromosome fusions and fissions, with substantial variation in genome size and male chiasmata localization. Historical cytological observations suggested that species with telocentric-rich karyotypes often retain interstitial male chiasmata, whereas predominantly biarmed karyotypes are frequently associated with strongly terminal male chiasmata. Yet this proposed relationship has not previously been evaluated across a broad phylogenetic sample. It also remains unresolved whether the strongly terminal male landscape represents an ancestral feature of anuran meiosis, a derived condition that increased in particular lineages, or a pattern repeatedly reshaped by karyotypic evolution.

Here, we synthesize published cytological and genetic-map data across amphibians, while accounting for phylogeny, to investigate the evolution of sex-specific recombination patterns. To distinguish variations in crossover localization from differences in overall recombination frequency, we consider three complementary aspects of recombination: the degree of terminal restriction of male recombination, spatial heterochiasmy between the sexes, and crossover rate heterochiasmy (table 1).

We first reconstruct the ancestral pattern of male CO location across the amphibian phylogeny. We then test whether variation in the terminal restriction of male recombination is associated with chromosome morphology, particularly centromere position, as well as chromosome number and genome size, and whether this pattern differs between XY and ZW systems. Finally, we integrate these comparative patterns to evaluate a framework in which an underlying sex-specific meiotic program interacts with chromosome morphology, especially centromere position, to generate variation in spatial heterochiasmy, and consider the broader consequences of this interaction for the evolution of heterochiasmy and sex chromosomes.

**Table 1. Distinct dimensions of recombination examined in this synthesis.** The degree of terminal restriction of male recombination, spatial heterochiasmy and crossover rate heterochiasmy describe complementary but distinct dimensions of meiotic recombination. Separating these measures allows variation in crossover localization to be distinguished from variation in overall recombination frequency.

| Recombination dimension | Definition | Measure | Coverage in this synthesis |
|---|---|---|---|
| Terminal restriction of male recombination | Degree to which male recombination is concentrated towards chromosome ends rather than distributed interstitially. | Relative prevalence of male bivalents bearing terminal versus interstitial chiasmata. | Broad comparative dataset encompassing 97 species. |
| Spatial heterochiasmy | Sex differences in the chromosomal localization of crossovers. | Comparison of male and female chiasma positions along chromosomes. | Subset of 20 species with comparable male and female CO localization data. |
| Crossover rate heterochiasmy | Sex differences in the overall frequency of meiotic crossovers. | Male–female differences in crossover number or genetic map length. | Subset of 9 species with sex-specific recombination measurements or genetic maps. |

## 2. Phylogenetic distribution and evolution of terminally restricted male recombination in amphibians

Historical comparative cytological studies revealed substantial variation in the locations of male chiasmata among amphibian lineages, establishing it as an evolutionarily variable feature rather than a uniform property of anuran meiosis. Morescalchi and Galgano (1973) [19] distinguished species in which large male bivalents retained multiple interstitial chiasmata from those in which chiasmata were predominantly restricted to terminal or subtelomeric regions. Several early diverging lineages, including Leiopelmatidae, Pipidae and Discoglossidae, contained large bivalents with multiple interstitial or subtelomeric chiasmata, whereas in many other anurans male bivalents typically had two chiasmata usually located near chromosome ends. On this basis, they proposed that the ancestral anuran condition involved more male interstitial recombination, followed by repeated evolution towards stronger terminally restricted recombination in multiple lineages [19].

Early cytological literature also suggested an association between centromere position and chiasma location. Species with predominantly telocentric chromosomes appeared more likely to retain interstitial male chiasmata, while chiasma position could also vary among chromosomes within the same karyotype [13,20]. By contrast, the more limited female cytological data generally show multiple chiasmata distributed broadly among chromosome arms, as observed on oocyte lampbrush chromosomes [21,22]. Modern linkage mapping has subsequently confirmed strongly terminally restricted male recombination across distantly related anuran families and across both XY and ZW sex-determining systems [14–18,23–25]. Together, these cytological and genetic-map studies show that male recombination localization varies substantially across amphibians, providing a basis for testing how this trait has evolved and whether its variation is associated with centromere position.

## A comparative framework for spatial heterochiasmy

To evaluate the evolutionary distribution of male chiasma location, we compiled published cytogenetic and genetic data for 97 amphibian species representing 20 anuran and 4 non-anuran families. We focus on male chiasma location because it is available across the broadest taxonomic samples and provides a measure that can be standardized across studies spanning many decades and methodologies. For each species, male chiasma location was quantified as the proportion of scored bivalents characterized by predominantly terminal or interstitial chiasmata. For nine species, male recombination pattern was inferred from sex-specific linkage maps. We additionally compiled haploid chromosome number, chromosome morphology, genome size, and sex chromosome system, when available.

The full dataset measures the degree of terminally restricted male recombination rather than heterochiasmy directly. Comparative female recombination information was available for only 20 of the 97 species, but in all cases female chiasmata were interstitial or broadly distributed along the chromosome arms. The smaller subset establishes pronounced terminally restricted male recombination in 16 species with data from both sexes, while four species have interstitial chiasmata in both males and females (figure 1; supplementary methods). The full dataset allows us to investigate the evolutionary history and chromosome-scale correlates of male chiasma location. Species-level classifications and data sources are provided in supplementary dataset S1.

## Evolutionary history of terminally restricted male recombination

The occurrence of predominantly interstitial male chiasmata in several early diverging amphibian lineages, together with the phylogenetically uneven distribution of terminally restricted male recombination, suggests that a more interstitial recombination pattern may present the ancestral condition. To investigate this possibility, we reconstructed the evolutionary history of male chiasma location using the species-level proportion of bivalents with terminal chiasmata and a time-calibrated composite amphibian phylogeny. The composite tree combined the broad amphibian phylogeny of Jetz & Pyron [26] , which included 7,238 species, with the more recent anuran phylogeny of Portik et al. [27], based on phylogenomic data and a large supermatrix comprising 5,242 frog species in a multi-loci supermatrix of Portik et al. [27].

Ancestral-state reconstruction inferred low proportions of terminal bivalents across the genome at deep amphibian nodes, including an estimate of 19.1% terminal bivalents for the ancestor of crown Anura (figure 1). However, uncertainty around these deeper estimates was substantial (supplementary figure S1; table S1; dataset S2). We therefore interpret the reconstruction as consistent with a predominantly interstitial ancestral condition rather than as definitive evidence for a particular ancestral state or a specific number of independent origins of terminally restricted recombination. Male chiasma location nevertheless showed clear phylogenetic signal (Pagel's $\lambda = 0.90$, $P < 0.001$; Blomberg's $K = 0.72$, $P < 0.001$), indicating that closely related species tend to resemble one another in the degree of terminal restriction of male recombination. The high $\lambda$ value indicates strong correspondence between trait variation and phylogenetic relatedness, whereas $K < 1$ suggests somewhat greater variation among close relatives than expected under a Brownian-motion model. Overall, these analyses indicate that terminal restriction of male recombination is evolutionarily labile but phylogenetically structured, with increased terminal CO localization arising in several anuran lineages. The variation in CO localization provides the basis for testing which chromosome features predict the degree of terminal restriction in male recombination.

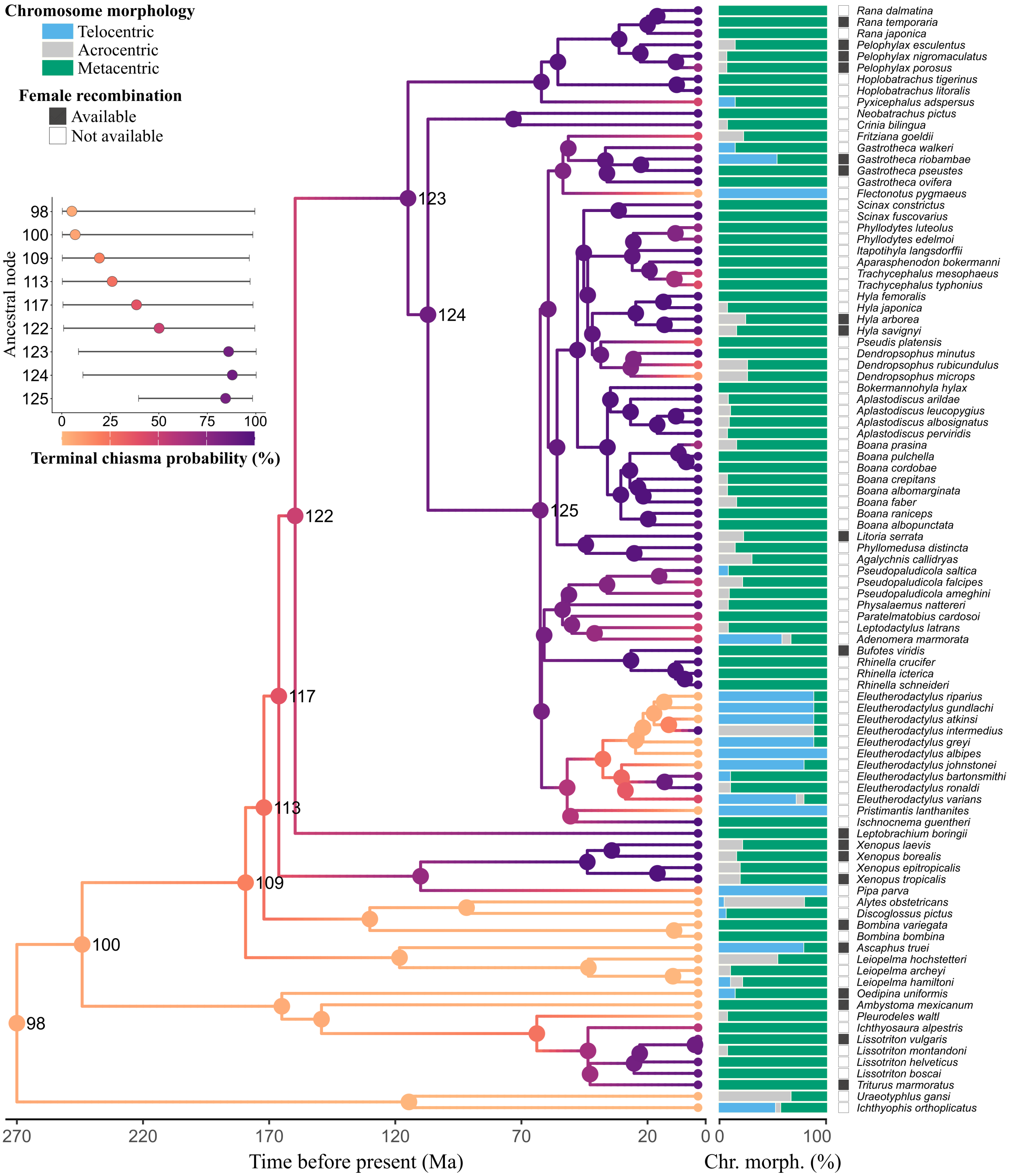

**Figure 1. Phylogenetic distribution of terminal restriction of male recombination across amphibians.** Tip nodes on the amphibian phylogeny represent the observed proportion of male bivalents with terminal chiasmata per species. Branch and internal-node colours represent reconstructed terminal chiasma probabilities, and numbered focal nodes correspond to the ancestral-state estimates shown in the inset, with horizontal bars indicating 95% confidence intervals. Tip-associated tracks summarize chromosome morphology as the proportions of telocentric, acrocentric and metacentric chromosomes. Squares indicate species for which comparable female recombination information is available. The reconstruction is consistent with relatively interstitial male recombination at deeper nodes and stronger terminal restriction in several anuran lineages, although uncertainty around deep ancestral states is substantial.

## 3. Chromosomal and genomic predictors of male crossover localization

Historical cytological studies show that species with predominantly telocentric chromosomes are more likely to retain interstitial male chiasmata [13,19,20], suggesting a possible role for centromere position in male crossover location. We therefore tested this after accounting for chromosome number and phylogenetic relatedness, using phylogenetic binomial generalized linear mixed models.

Centromere position showed the strongest association with male chiasma location (figure 2; supplementary material, table S2a). Species with a higher proportion of telocentric chromosomes tended to have a lower proportion of male bivalents with terminal chiasmata ($P = 5.4 \times 10^{-8}$), whereas species with a higher proportion of metacentric or submetacentric chromosomes showed stronger terminal localization ($P = 1.3 \times 10^{-5}$). Acrocentric and subtelocentric chromosome prevalence showed a weaker positive association with terminal localization ($P = 0.009$; supplementary material, table S2a), while chromosome number showed no consistent association with CO location across models (supplementary material, table S2a).

Because chromosome-morphology proportions are compositional, these associations should be interpreted collectively rather than as independent effects of individual chromosome classes. Nevertheless, phylogenetically controlled models revealed pronounced and contrasting relationships between chromosome morphology and male chiasma location. Increasing telocentric chromosome prevalence was associated with a marked decline in the predicted proportion of terminal male bivalents, from ~6% at the mean prevalence to <1% at higher prevalence, whereas increasing metacentric/submetacentric prevalence showed the opposite pattern, with terminal bivalents increasing from ~8% to ~43% across the biologically valid modelled range (supplementary material, table S2b). Acrocentric/subtelocentric chromosomes showed a weaker positive association, with predicted terminal bivalents increasing from ~8% to ~16% (supplementary material, table S2b). Thus, telocentric-rich karyotypes tend to retain more interstitial male chiasmata, whereas karyotypes enriched for chromosomes with internally positioned centromeres show stronger terminal restriction. Although absolute predictions remain uncertain, these relationships persist after accounting for phylogenetic relatedness.

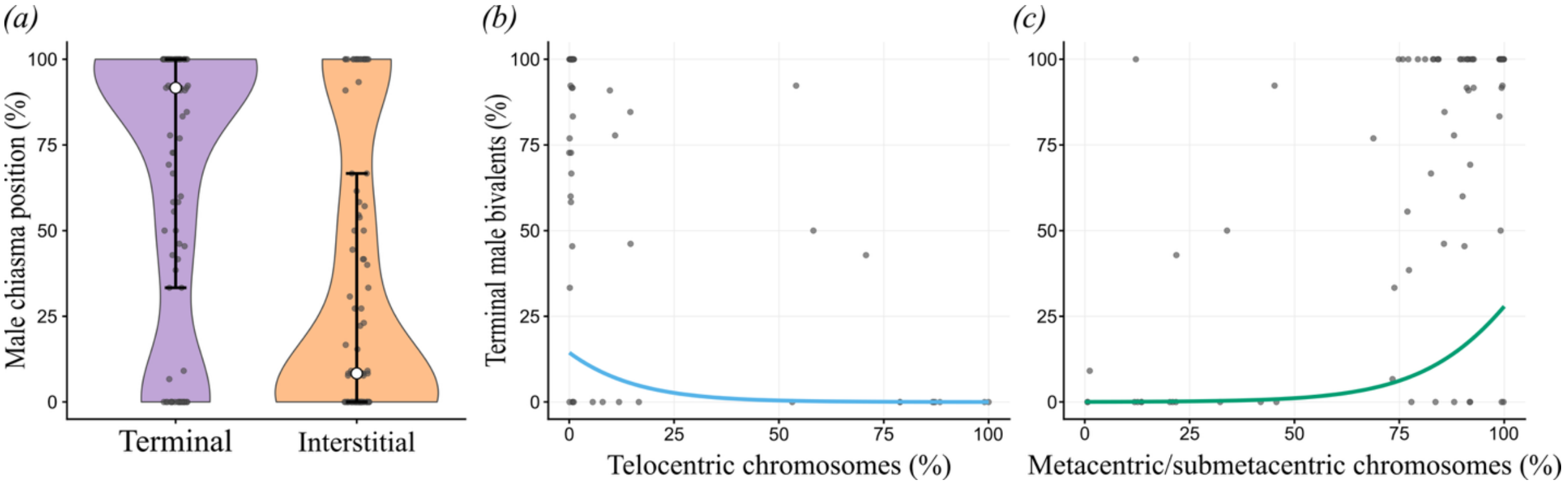

**Figure 2. Association between male chiasma location and centromere position across amphibians.** (a) Species-level distribution of the proportions of male bivalents with terminal and interstitial chiasmata across 97 species. Points represent individual species; white circles indicate medians and black bars indicate interquartile ranges. (b-c)

Relationship between the proportion of male bivalents with terminal chiasmata and the prevalence of (b) telocentric and (c) metacentric/submetacentric chromosomes. Solid lines show fitted relationships from separate phylogenetic binomial generalized linear mixed models, with standardized log bivalent number held at its mean and phylogenetic random effects set to zero; points represent individual species. Telocentric-rich karyotypes are associated with a lower proportion of terminal male bivalents, whereas karyotypes with a greater prevalence of metacentric/submetacentric chromosomes show stronger terminal chiasma localization.

### ***Eleutherodactylus* as a within-lineage illustration**

The genus *Eleutherodactylus* provides a particularly informative comparative example because both centromere position and male chiasma location vary extensively among related species. Cytogenetic data from more than 200 terraranan frog species show extensive karyotype diversity, including recurrent centric fusions and fissions, variation in centromere position, and pronounced differences in chromosome number and morphology, with diploid numbers ranging from approximately 2n = 16-38 [13].

Within *Eleutherodactylus*, species with strongly telocentric karyotypes (*E. albipes, E. atkinsi, E. greyi, E. gundlachi, E. johnstonei* and *E. riparius*), generally have interstitial male chiasmata (figure 1). In contrast, species with few or no telocentric chromosomes, (*E. intermedius* and *E. ronaldi*, *E. bartonsmithi*) showed predominantly terminal male chiasmata. *E. varians* was intermediate in both chromosome morphology and chiasma location. These observations do not constitute an independent test of causality, but they provide a useful within-lineage example of the broader comparative association between centromere position and male crossover position.

The association is consistent with a model in which centromere position modifies how an underlying male-specific crossover placement is expressed along chromosomes. CO formation is constrained by crossover assurance and interference, while recombination is generally suppressed in centromeric and pericentromeric regions; moreover, the relevant unit of crossover assurance can vary between whole chromosomes and individual chromosome arms [4,28]. Experimental and comparative studies in other taxa similarly show that chromosome size and structure help shape broad-scale recombination landscapes, with centromeric regions typically constraining local CO formation [29]. Thus, changes in centromere position or arm configuration could alter the spatial expression of a male-specific recombination program without requiring that program itself to change.

In a male recombination landscape already biased toward chromosome ends, a biarmed chromosome divides the chromosome axis into two shorter arms separated by a centromere-proximal region of reduced recombination. By contrast, a telocentric chromosome provides one long arm over which additional interstitial COs may be accommodated. This offers a plausible explanation for the comparative association, but centromere position alone cannot generate heterochiasmy because it is shared by both sexes. Rather, it must modify the expression of sex-specific regulation of the meiotic recombination machinery as we will discuss in Section 4.

### **Limited evidence for effects of chromosome or genome size**

We next asked whether genome size and mean chromosome size can explain additional variation in male chiasma location beyond centromere position. These predictors were tested in the 54 species for which genome-size data were available, while accounting for chromosome morphology, haploid chromosome number and phylogenetic relatedness. In these models, centromere position remained significantly associated with male chiasma location. Telocentric chromosome prevalence retained a strong negative association with terminal location ($P = 6.3 \times 10^{-7}$), whereas metacentric/submetacentric prevalence remained positively associated (*P* = 0.0075). By contrast, neither genome size nor mean chromosome size showed independent statistical support (*P* = 0.082 and 0.194, respectively, depending on the chromosome-morphology model; supplementary material figure S2; table S3).

Mean chromosome size should be treated cautiously as a proxy for meiotic chromosome length. During meiotic prophase, chromatin is organized into loops anchored to proteinaceous chromosome axes; as homologues synapse, these axes become connected along their lengths by the synaptonemal complex (SC). Recombination is tightly integrated with this SC–axis organization, and CO number is more closely associated with meiotic axis or SC length than with physical DNA content alone [30]. We therefore regard chromosome- and axis-length effects as mechanistic hypotheses requiring direct cytological measurement rather than conclusions supported by the present comparative analyses.

It is important to distinguish absolute CO number from recombination rate per unit of DNA length. Larger chromosomes can carry more COs in absolute terms. For example, CO number increases with both physical chromosome size and SC length in *Arabidopsis*, and comparative analyses across flowering plants likewise show that linkage-map length is influenced by chromosome size [29,31]. However, this increase is typically sublinear, so smaller chromosomes tend to have higher recombination rates per megabase, partly reflecting the requirement for at least one CO per bivalent. Consistent with this distinction, Zedek et al. [32] found that genome-wide recombination rate decreases with increasing mean chromosome size across insects, with a similar relationship in monocentric flowering plants. Thus, chromosome size can influence the overall amount or density of recombination without necessarily determining its spatial distribution. This distinction is important for our amphibian analyses, in which mean chromosome size showed no independent association with male chiasma location after chromosome morphology was accounted for.

The spatial placement of COs may depend on both meiotic chromosome organization and sex-specific recombination processing. Differences in meiotic axis or SC length could contribute to heterochiasmy by altering the physical context in which COs are formed and spaced, but axis length alone is unlikely to explain sex-specific patterns. In mice, CO number and distribution can differ between the sexes independently of SC length, reflecting sex-specific differences in CO designation, interference and pathway usage [33]. Direct measurements of meiotic axis and SC length, together with recombination-processing markers, will therefore be needed to resolve the structural and regulatory basis of spatial heterochiasmy.

**Male chiasma location is largely independent of heterogamety**

Finally, we tested whether male chiasma location differs between male-heterogametic (XY) and female-heterogametic (ZW) systems. This comparison is particularly relevant because amphibians show frequent sex chromosome turnover, including transitions that either preserve or switch the heterogametic sex. In true frogs, most inferred turnovers retain male heterogamety, whereas in *Xenopus* most known turnovers retain female heterogamety, although switches between XY and ZW systems have occurred in both lineages [18,34]. These repeated transitions provide a natural context in which to ask whether male recombination localization is associated with the heterogametic sex or instead represents a more conserved feature of male meiosis.

Among the 40 species in our dataset with characterized sex chromosome systems (XY vs ZW), we found no statistically supported association between heterogamety and male chiasma location (*P* = 0.072) after accounting for chromosome morphology, chromosome number and phylogenetic relatedness (supplementary material figure S3; table S4). The result should not be interpreted as evidence that XY and ZW taxa have equivalent male CO locations, because the analysis had limited power both because of the small number of species with data, and because it was strongly biased toward XY species (i.e. 28 XY vs 12 ZW species). However, the analysis provides no current support for heterogametic system as the primary explanation for the genome-wide male recombination pattern. This interpretation is consistent with the strongly terminal male recombination observed in the ZW *Hyla* species despite at least 11 million years of female heterogamety [17].

Taken together, the comparative analyses narrow the explanatory framework. Male terminal chiasma localization is phylogenetically structured and strongly associated with metacentric/submetacentric chromosome morphology (supplementary material, table S2a), whereas we detected no significant association with haploid chromosome number, genome size, mean chromosome size, or heterogametic system; however, the latter analyses were based on more limited datasets (supplementary material, table S3, S4). Centromere position alone, however, cannot explain chiasma location differences between the sexes. Spatial heterochiasmy must therefore involve a sex-specific meiotic component, with centromere position potentially modifying how strongly underlying differences in chiasma location are expressed. The next section considers which stages and processes of meiotic prophase may generate these sex-specific recombination patterns.

## 4. From comparative pattern to meiotic mechanism

The first question regarding the mechanism of spatial heterochiasmy is whether it is specified by genotypic sex or by the meiotic environment associated with phenotypic sex (figure 3a). Sex-reversed individuals, i.e. individuals with the genotype associated with one sex that develop as the other sex, such as XY females [35,36], provide a direct test of these alternatives. Evidence from sex-reversal studies in frogs, fishes and non-avian reptiles strongly supports the view that sex-specific meiotic environment follows phenotypic rather than genotypic sex [37–39]. For example, recombination pattern follows phenotypic sex rather than sex chromosome genotype in *Rana temporaria*, where naturally occurring XY females recombine as extensively as XX females, with sex chromosome map lengths of approximately 150 cM, whereas XX males show the strongly restricted recombination characteristic of XY males, with average map lengths of only about 2 cM [37]. A strikingly similar pattern has recently been reported in rainbow trout, where hormonally sex-reversed XX males showed the subtelomeric CO distribution and CO number characteristic of XY males rather than XX females [39].

We interpret the data to support spatial heterochiasmy arising from sex-specific regulation of meiosis acting on a shared chromosome architecture between the sexes. We now turn our focus to meiotic prophase stages to investigate in which aspect of the shared recombination machinery between the sexes works in a sex-specific fashion to cause a sex-specific CO localization.

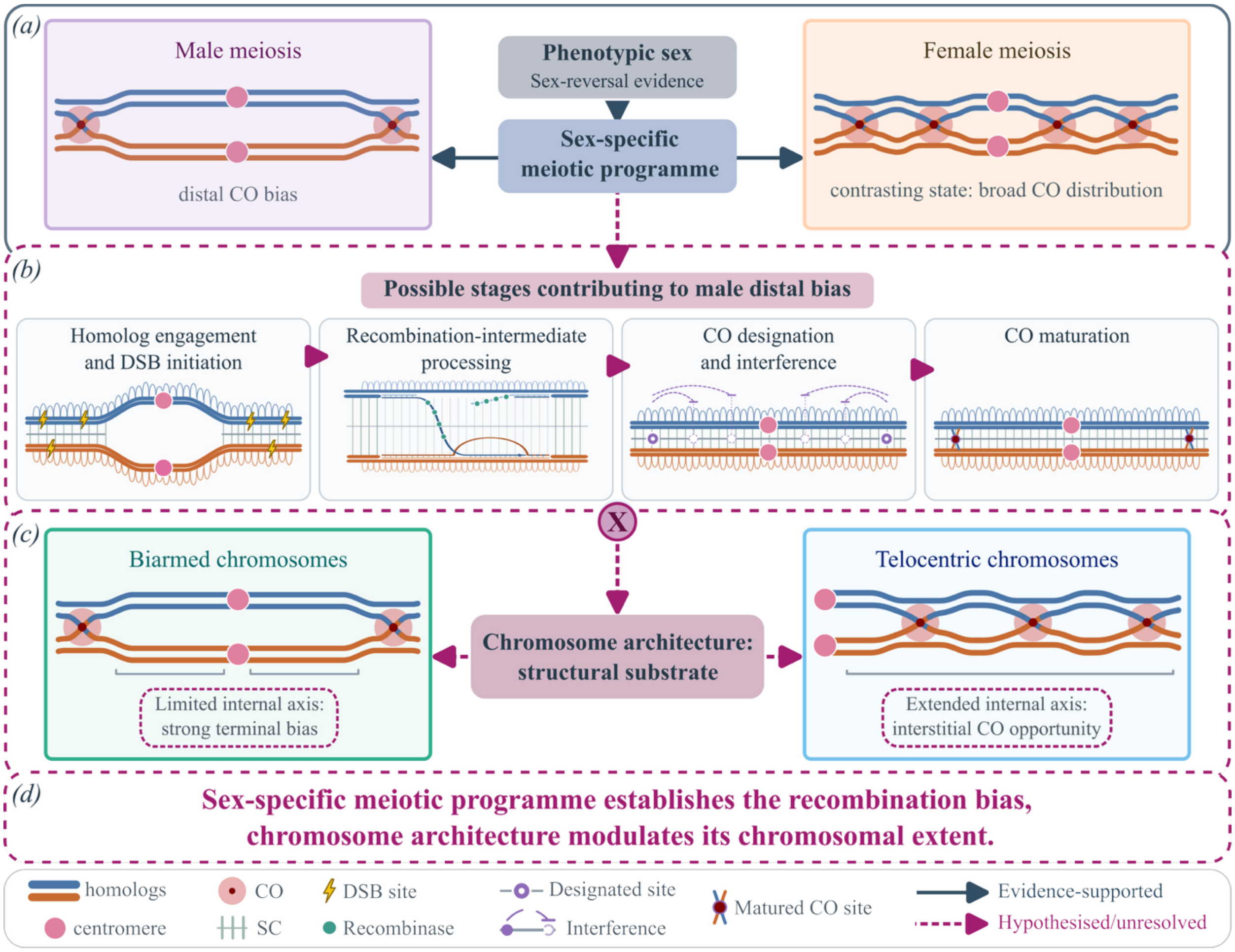

**Figure 3. A sequential meiotic-program × chromosome-architecture model for extreme spatial heterochiasmy.** (a) Sex-reversal evidence indicates that sex-specific recombination patterns largely follow phenotypic rather than genotypic sex, supporting a major role for the sex-specific meiotic program associated with male vs female meiosis. (b) A distal male CO bias could arise at one or more stages of meiotic recombination, including homolog engagement and DSB initiation, processing of recombination intermediates, CO designation and interference, and CO maturation. These stages are presented as alternative or interacting possibilities rather than as a single established pathway. (c) Chromosome architecture provides the structural substrate on which this sex-specific meiotic program operates. Under the same male program, biarmed chromosomes may show stronger terminal restriction because distal COs, CO interference and centromere-associated suppression leave relatively little internal axis available for additional COs, whereas telocentric chromosomes may retain a larger internal region in which interstitial COs can occur. (d) The model therefore proposes that a sex-specific meiotic program establishes the recombination bias, while chromosome architecture modulates its chromosomal extent. Solid arrows indicate relationships supported by available evidence; dashed arrows and boxes indicate hypothesized or unresolved mechanistic relationships. CO, crossover; SC, synaptonemal complex; DSB, double-strand break.

## Where is the spatial bias established?

Spatial heterochiasmy could arise at several stages between homolog chromosome engagement and mature CO formation. During meiotic prophase, chromosomes first organize into loop-axis structures that provide the framework for DSB formation, homolog pairing and SC assembly. Recombination then proceeds through DSB repair and intermediate processing to CO designation, patterning and maturation. Sex differences have been documented at several of these stages in model systems (Cahoon & Libuda, 2019; Ishiguro, 2022). Importantly, these stages need not contribute equally to spatial heterochiasmy: where recombination is initiated does not necessarily determine where mature COs ultimately occur.

The limited available amphibian cytology suggests that chromosome ends may have an early spatial advantage during homolog pairing and recombination. In *Xenopus laevis*, SC ends become markedly thickened and cluster at the nuclear periphery during zygotene [40], a bouquet-like configuration consistent with telomere-associated organization during early homolog engagement. In male *Leptodactylus pentadactylus*, synapsis initiates between homologous terminal regions, while γ-H2AX is strongly enriched at chromosome ends and RAD51 frequently associates with telomeric regions during zygotene [41]. Because γ-H2AX marks chromatin associated with meiotic DSB formation and repair, whereas RAD51 promotes homology search and strand invasion during meiotic DSB repair, their distal enrichment indicates preferential localization of early recombination activity near chromosome ends. This observation should nevertheless be interpreted cautiously, because *L. pentadactylus* carries an exceptional multivalent sex-chromosome system in which terminal homology contributes directly to chromosome pairing. Together, these studies make distal-first homolog engagement and early recombination processing plausible in amphibians, but broader cytological evidence is needed to determine how general this pattern is and whether it contributes causally to terminally restricted male recombination in amphibians.

Importantly, studies in other organisms suggest that the final CO landscape can diverge strongly from the initial distribution of recombination intermediates. In mice, Brick et al. [42] showed extensive sex differences in DSB hotspot activity, but also demonstrated sex-specific differences in the probability that distal DSBs ultimately become COs. More recent work across genetically divergent mouse strains further shows that SC length alone cannot explain sex differences in CO number: the efficiencies of later recombination-processing steps, including CO designation, are themselves sexually dimorphic [33]. In humans, there is sexual dimorphism in the probability of maturation of CO-designated intermediates into final COs [43]. Spatial heterochiasmy therefore depends not only on where recombination is initiated, but also on sex-specific processing of recombination intermediates that determines where mature COs ultimately form.

The SC is closely involved in these processes particularly through CO interference and assurance. In *Arabidopsis thaliana*, loss of the SC transverse-filament protein ZYP1 abolishes CO interference and simultaneously erases the normal difference in CO number between male and female meiosis, largely through an increase in female COs [44]. Similarly, loss of the SC central-element protein SCEP3 disrupts both CO assurance and interference and eliminates heterochiasmy, through disproportionately increasing female COs [45]. These results show that SC-dependent regulation of CO designation and spacing can generate sex-specific recombination differences. Axis-SC organization, interference and CO maturation therefore provide plausible mechanisms through which sex-specific processing could contribute to spatial heterochiasmy in amphibians.

## A sequential model of interaction between the sex-specific meiotic program and chromosome architecture

We propose that extreme spatial heterochiasmy in amphibians can emerge through an interaction between centromere position and multiple stages of a sex-specific meiotic program (figure 3). Under this model, spatial heterochiasmy is specified by the meiotic environment associated with phenotypic sex (figure 3a). Distal chromosome regions may gain an early advantage during homolog engagement and recombination initiation, while sex differences may arise later through preferential crossover designation or maturation near chromosome ends in males (figure 3b). CO interference could then constrain the placement of additional COs once designation has begun (figure 3c), reinforcing terminally restricted male recombination. The extent of this restriction may subsequently depend on chromosome architecture, as discussed below.

Chromosome architecture could then determine how strongly this underlying male bias is expressed (figure 3c). On biarmed (metacentric/submetacentric) chromosomes, an internally positioned centromere divides the chromosome into two shorter arms and creates a central region in which CO formation is reduced. If male COs are already biased towards chromosome ends, CO interference around distal events combined with centromeric suppression could strongly reduce the probability of additional COs forming in internal chromosome regions [24,46]. This would reinforce terminal chiasma location and could contribute to the characteristic ring-shaped bivalents frequently observed in male frogs. In females, longer chromosome axes and/or a greater probability that interstitial recombination intermediates mature as COs could allow recombination to extend farther into chromosome interiors. A telocentric chromosome instead provides one relatively long arm allowing for larger internal segments that remain outside both the CO interference domain of any terminal COs and the centromere-proximal region of reduced recombination, increasing the opportunity for interstitial CO formation. This provides a mechanistic interpretation of the strong comparative association between telocentric-rich karyotypes and retention of interstitial male chiasmata.

Under this framework, sex-specific meiosis establishes the spatial bias, while centromere position determines the chromosomal extent of that bias. This distinction also explains why changes in chromosome architecture could alter male recombination landscapes without requiring repeated evolution of the underlying molecular machinery. Centromere repositioning, chromosome fusion or fission could change the physical relationship among chromosome ends, centromeres and interference domains and thereby strengthen or weaken terminal restriction.

Importantly, this model does not require DSB formation to be restricted to chromosome ends, nor does it require shorter male chromosome axes to be the universal cause of heterochiasmy. Both remain testable possibilities. The central prediction is instead that spatial heterochiasmy should become progressively detectable between recombination initiation and mature CO formation, and that its final spatial expression should depend on chromosome-arm configuration.

**Testable predictions**

1. If terminal restriction of male COs originates during DSB initiation, DSB sites should already show a stronger distal bias in males than females. If it emerges during subsequent recombination-intermediate processing, markers of early recombination intermediates should show progressively stronger distal enrichment relative to DSB sites. If it arises primarily during CO designation or maturation, DSBs and early intermediate markers should remain more broadly distributed than mature male COs, with terminal bias becoming more pronounced as the recombination pathway proceeds.

2. Males and females should differ in chromosome-axis organization, CO interference, or the position-dependent probability that recombination intermediates are designated and matured as COs, and the magnitude of these differences should covary with the intensity of spatial heterochiasmy across species.

3. After accounting for meiotic axis length, telocentric chromosomes should retain more interstitial male COs than biarmed chromosomes; independently, longer chromosome axes should increase the opportunity for additional interstitial COs.

4. In sex-reversed individuals, the localization of COs and earlier recombination intermediates should follow phenotypic rather than genotypic sex.

These predictions can be tested through cytological comparisons of male and female meiosis and, complemented where feasible by direct DSB mapping and 3D genome analyses. Meiotic chromosome spreads with immunofluorescent labelling could quantify chromosome axes and SC organization (e.g. SYCP3/SYCP1), early recombination intermediates (RAD51/DMC1), and class I CO designation and maturation (e.g. HEI10/MLH1) [30,47]. Comparing marker distribution, together with axis length and CO spacing, would reveal when the male terminal bias emerges along the recombination pathway. Direct DSB mapping, for example using SPO11-oligo or related approaches, could distinguish biases established at recombination initiation from those arising during subsequent processing [48], while Hi-C or related chromosome-conformation approaches could test whether sex-specific meiotic chromosome organization contributes to the pattern [49]. Applying these comparisons across sexes, chromosomes and, where available, sex-reversed individuals would directly test the proposed interaction between the sex-specific meiotic programme and chromosome architecture.

## 5. Why does extreme spatial heterochiasmy evolve?

Understanding the mechanism of terminal male recombination does not explain why such a recombination landscape evolved. The distinction between proximate mechanism and ultimate evolutionary cause is particularly important for heterochiasmy because several non-exclusive hypotheses have been proposed, yet broad comparative analyses across eukaryotes have not identified a universal explanation [7]. Our amphibian-focused comparative analyses provide a more detailed test of these ideas by incorporating phylogenetic relationships and variation in chromosome morphology, heterogamety and other genomic features, allowing existing hypotheses to be evaluated in a clade characterized by particularly extreme spatial heterochiasmy.

### Mechanistic constraint and indirect evolution through chromosome architecture

The simplest interpretation of the amphibian data is that the proximate mechanism of extreme spatial heterochiasmy is a sex-specific meiotic program. Terminally restricted male recombination across most autosomes (i.e. >75%) in more than 60 of the 97 species examined, is found under both XY and ZW systems. Independent sex-reversal studies further show that the same chromosome can experience profoundly different recombination regimes depending on whether it passes through male or female meiosis [36,37,39]. Together, these analyses suggest that the genome-wide male recombination pattern is not generated by sex chromosome differentiation; rather, sex chromosome differentiation may arise secondarily when a chromosome carrying a sex-determining locus is repeatedly exposed to the male meiotic program.

However, the pattern could simultaneously have an adaptive component. Spatial heterochiasmy itself may have evolved through sex-specific selection on the meiotic program, while its present chromosomal expression has subsequently been modified by karyotypic evolution, including changes in centromere position and centric fusions or fissions. Such architectural changes could alter how an existing male recombination program is expressed along chromosomes, although these changes need not be selectively neutral and may themselves be shaped by processes such as meiotic drive or selection on chromosome segregation. Thus, selection may act both on the meiotic program and on chromosome architecture, but not necessarily for the same reasons. Distinguishing the evolution of the sex-specific meiotic program from the evolution of the chromosomal context in which it is expressed is therefore an important implication of the amphibian comparison.

**Adaptive hypotheses**

One adaptive explanation for heterochiasmy invokes differences in selection during the haploid phase. Theory proposes that heterochiasmy can evolve when males and females experience different patterns of haploid or parent-of-origin-dependent selection, because recombination changes the combinations of alleles exposed to selection [50]. Comparative analyses in plants subsequently found that sex differences in recombination covary with the relative opportunity for haploid selection in male and female gametophytes: the sex experiencing stronger haploid selection tends to show lower recombination [8]. The relevance of this mechanism to amphibians is untested. Sperm can experience selection before fertilization, but the extent to which haploid gene expression and epistatic selection are sufficiently strong to generate chromosome-wide terminal recombination in amphibians is unknown. Moreover, the hypothesis was developed primarily to explain differences in overall recombination rate, not the repeated evolution of an extreme terminal landscape. It remains a plausible ultimate explanation for spatial heterochiasmy, but current amphibian data lack the comparative estimates of haploid selection needed to test it.

Female meiotic drive provides another potential source of sex-specific selection on recombination. Because female meiosis is asymmetric, centromere variants that bias their transmission to the functional egg have an evolutionary advantage. This can cause the counter-evolution of variation that modifies this transmission advantage, such as CO position modifiers [51]. This mechanism provides a compelling reason for more recombination close to centromeres in oogenesis, but does not explain why male COs should concentrate in chromosome ends. An extreme sex difference may nevertheless not only result from forces operating during male meiosis; selection on the female landscape could also increase sexual dimorphism. At present, the amphibian data do not allow these alternatives to be distinguished.

A third class of adaptive hypotheses links heterochiasmy to sex differences in selection on multilocus genetic associations. Trivers (1988) [52] proposed that the sex experiencing greater variance in reproductive success should evolve lower recombination, because recombination can disrupt favorable combinations of alleles associated with reproductive success. This predicts particularly reduced male recombination in species subject to strong sexual selection on males. Empirical support, however, has been mixed. Across eutherian mammals, Sexual dimorphism was associated with heterochiasmy, but in the opposite direction to the Trivers prediction: male recombination was higher, rather than lower, in more sexually dimorphic species [6]. Similarly, a comparative analysis of 61 fish species found rapid evolutionary shifts in heterochiasmy but no significant relationship with proxies for sexual selection or sexual conflict [53]. Thus, sex-specific selection on genetic associations remains a plausible contributor to heterochiasmy, but current comparative evidence does not support it as a general explanation for sex-specific recombination patterns.

**Selection on chromosome segregation and the boundary with achiasmy**

CO number and position also affect the mechanical reliability of chromosome segregation, making sex-specific selection against aneuploidy another potential force. This possibility is particularly relevant because the relationship between recombination and segregation differs between spermatogenesis and oogenesis, and because the consequences of poorly positioned COs can strongly depend on sex [7,43]. At the extreme, recent theory on achiasmatic meiosis suggests that different selective forces may dominate at different stages of sex chromosome differentiation, with sexually antagonistic selection being most important when sex chromosomes are young and selection against segregation errors becoming more important as heteromorphy increases [54]. That model concerns achiasmy rather than amphibian spatial heterochiasmy directly, but it illustrates that the evolutionary cause of recombination suppression may change through time.

The current amphibian data cannot distinguish among these ultimate hypotheses. What they do show is that heterogamety itself is insufficient to explain variation in male CO position, whereas centromere position has a much stronger association with it. This shifts the evolutionary problem away from searching for a single explanation involving the sex chromosomes and toward understanding how selection and constraint act on the male and female meiotic programs and their influence by centromere position.

The selective forces that originally produced a sex-specific recombination program need not be the same as those that subsequently shape or maintain it. Once sex differences in meiotic axis organization, CO designation or interference have evolved, they may become integrated with other aspects of gametogenesis and constrained by pleiotropy. Subsequent evolution of chromosome architecture, including changes in centromere position or chromosome-arm configuration, could then modify the chromosomal expression of that program. Such changes need not be selectively neutral: centromere position and chromosomal rearrangements may themselves evolve under selection, including meiotic drive, and may also be favored or constrained through their effects on recombination and chromosome segregation. Selection could likewise continue to modify the meiotic program itself in particular lineages. Extreme spatial heterochiasmy may therefore reflect the combined effects of historical selection, ongoing selection on meiotic and chromosomal traits, and mechanistic constraint, rather than the action of a single selective pressure.

**BOX: Convergent extreme spatial heterochiasmy in teleost fishes**

Teleost fishes provide an informative comparison because a strong distal bias in male recombination has evolved in multiple lineages, but is not a universal feature of teleost recombination landscapes. For example, male recombination concentrated near chromosome ends and more broadly distributed female recombination have been documented across several salmonid lineages, including brown trout, brook charr and Atlantic salmon, despite substantial variation among species and chromosomes in the magnitude of sex differences in total recombination [55–58].

Two examples are especially informative. In brook charr (*Salvelinus fontinalis*), females have approximately 2.7 times as many COs as males, while male COs occur mainly within the distal portions (outer 20%) of linkage groups. Spatial heterochiasmy is also stronger on fused metacentric than acrocentric chromosomes, indicating as in frogs that chromosome architecture may affect sex-specific recombination, although this comparison is partly confounded by residual tetraploidy and incomplete male maps [57]. In atlantic salmon (*Salmo salar*), the female linkage map is only approximately 1.47-fold longer than the male map, but the predominantly subtelomeric distribution of male COs produces approximately 8-fold less intrachromosomal genetic shuffling [58]. This illustrates why crossover position can have much greater evolutionary consequences than differences in overall recombination rate.

Terminal male recombination occurs in other teleosts, including Brown trout (*Salmo trutta*), rainbow trout (*Oncorhynchus mykiss*), Chinook salmon (*Oncorhynchus tshawytscha*), the Trinidadian guppy (*Poecilia reticulata*), and zebrafish (*Danio rerio*) [39,55,56,59,60]. In contrast, *Xiphophorus* shows little evidence of sex-specific differences in recombination, whereas in medaka recombination suppression is largely localized around the sex-determining region [60–62]. Together with the pronounced sex differences in CO distribution observed in other teleosts, these examples indicate substantial evolutionary variation in the magnitude and spatial distribution of heterochiasmy across teleost lineages, rather than a single conserved teleost pattern. In salmonids, whole-genome duplication and the persistence of residual homology between duplicated chromosome arms may have contributed to the evolution of terminally localized COs because they could facilitate chromosome pairing and segregation while limiting exchange among residually duplicated regions during male meiosis [57,63]. However, these lineage-specific features cannot explain similar patterns in distantly related teleosts. Collectively, teleosts support a model in which sex-specific meiotic regulation interacts with chromosome architecture and lineage history to generate convergent extreme spatial heterochiasmy.

# 6. Genomic and sex chromosome consequences of extreme spatial heterochiasmy

A tendency for CO close to a telomere satisfies the mechanical requirement for homolog segregation while leaving most of the chromosome inherited as a large, linked block. The resulting extreme spatial heterochiasmy is evolutionarily important because CO rate alone is insufficient to characterize chromosome reshuffling. The size and distribution of chromosome parts that are reshuffled are responsible for linkage disequilibrium, linked selection and the evolutionary independence of loci across the genome. Nevertheless, even recombination in one sex is sufficient to break genetic associations, so that extreme spatial heterochiasmy does not lead to complete recombination suppression. A notable exception are alleles preferentially transmitted through males or females, for example the alleles on X and Y chromosome: they experience different opportunities for reshuffling than autosomal alleles, which has big consequences for sex chromosome evolution. Sardell and Kirkpatrick (2020) [7] similarly emphasized that sex-specific recombination landscapes can have broader evolutionary consequences, including effects on patterns of population differentiation as well as on sex chromosome evolution.

**Pre-existing recombination landscapes can shape sex chromosome evolution**

Classical models of sex chromosome evolution usually begin with the origin of a sex-determining locus followed by selection for reduced recombination around it, potentially due to sexually antagonistic alleles and structural rearrangements [64,65]. This causal sequence may not be universal. If a new sex-determining locus arises within a genomic region that already experiences very little recombination in the heterogametic sex, extensive sex linkage would arise without requiring the evolution of an additional recombination-suppression mechanism. Evidence from *Rumex hastatulus* illustrates this reverse causal direction particularly clearly. Rifkin et al. (2022) [11] showed that both ancestral and neo-sex-linked regions are embedded within broad regions of low recombination and that low recombination largely preceded acquisition of sex linkage. Thus, the ancestral recombination landscape facilitated sex-chromosome differentiation rather than simply evolved as its consequence.

Amphibians provide compelling parallels. In *Leptobrachium boringii*, X–Y differentiation extends across approximately 349 Mb, yet there is little evidence that chromosomal inversions generated this exceptionally large sex-linked region. Instead, the differentiated region largely overlaps a region of ancestrally low male recombination, suggesting that the pre-existing recombination landscape allowed X–Y differentiation to extend across much of the chromosome after the male-determining factor arose [25]. This differentiation is nevertheless counteracted by sex reversal: 8 of 97 phenotypic females (8.2%) carried an XY genotype, allowing X–Y recombination during female meiosis and potentially helping to maintain overall X–Y homomorphy despite the extensive differentiated region. A related mechanism has been proposed in Ranidae. Because male recombination is strongly restricted toward chromosome ends, recruitment of a chromosome for male sex determination can cause most of that chromosome to become effectively non-recombining between X and Y immediately upon the establishment of sex linkage [34,66]. This represents a departure from the canonical model in which recombination suppression spreads progressively outward from a newly established sex-determining region: under extreme spatial heterochiasmy, extensive X–Y differentiation can instead begin across much of the chromosome from the onset of sex linkage. Jeffries et al. (2018) [34] further proposed that the resulting rapid accumulation of mutation load on the Y could contribute to the exceptionally high rate of sex-chromosome turnover observed in ranid frogs.

However, the recombination landscape need not determine where on a chromosome a new sex-determining region originates. *Xenopus* provides an informative counterexample to the hypothesis that new sex-determining regions preferentially arise in ancestrally low-recombination regions. Across the genus, Evans et al. (2024) [18] identified at least eight non-homologous sex-associated regions and found that these regions occur significantly closer to chromosome ends than expected by chance, despite chromosome ends having relatively high ancestral recombination rates. Thus, the locations at which sex linkage originates do not appear to be preferentially drawn from ancestrally recombination-suppressed regions in *Xenopus*. After sex linkage arose, however, sex-associated recombination arrest expanded in several lineages, although whether this expansion was driven directly by heterochiasmy or by other mechanisms remains unresolved.

Together, *Leptobrachium*, ranids, and *Xenopus* suggest that two effects of recombination landscapes should be distinguished. Pre-existing recombination patterns do not universally predict where along a chromosome a new sex-determining region will originate, but they can strongly influence how much of the chromosome becomes effectively sex linked and, consequently, the subsequent trajectory of X–Y or Z–W differentiation.

**Sex reversal can reverse the consequences of recombination suppression**

Extreme heterochiasmy also creates an unusual route by which sex reversal can restore recombination between sex chromosomes [36]. Perrin's (2009) [35] "fountain-of-youth" model proposed that if recombination is determined by phenotypic rather than genotypic sex, occasional XY females can allow X–Y recombination and thereby rejuvenate a degenerated Y-chromosome. Rodrigues et al. (2018) [37] subsequently provided field evidence for this in *R. temporaria*: XY females recombined extensively, whereas XX males showed male-like recombination suppression. The recent *L. boringii* study extends this idea to a very large sex-linked region. Xie et al. (2025) [25] found a substantial frequency of sex-reversed XY females and argued that recombination during female meiosis can limit X–Y sequence divergence and degeneration across the large sex-linked region. Thus, although extreme male heterochiasmy promotes extensive sex linkage, occasional passage of the Y through female meiosis can restore X–Y genetic exchange and counteract further differentiation.

These observations reveal feedback among heterochiasmy, sex reversal and sex chromosome turnover. Extreme terminal restriction of male recombination can accelerate differentiation of newly recruited Y chromosomes; mutation load may subsequently favour turnover, as proposed for ranids; alternatively, occasional sex reversal can restore X–Y recombination and maintain homomorphy. The evolutionary fate of amphibian sex chromosomes may therefore depend not simply on whether recombination is suppressed, but on where recombination occurs, which phenotypic sex controls that landscape, and how chromosome architecture constrains its physical extent.

## 7. Synthesis and testable predictions

The defining characteristic of extreme spatial heterochiasmy is the difference in CO position between the sexes. It is usually more extreme than CO rate heterochiasmy, and has more profound evolutionary consequences. Our comparative synthesis further shows that the degree of male terminal chiasma localization is not fixed, but is strongly associated with chromosome architecture. We propose a two-level model in which a sex-specific meiotic program establishes a bias in CO-precursor placement, while centromere position influences which precursors mature into COs and thereby shapes the chromosome-scale recombination landscape.

Evidence from species with sex reversal indicates that CO precursor placement largely follows phenotypic sex, whereas the phylogeny-independent association of CO location with centromere position suggests that the CO location secondary evolves with karyotype architecture. Consequently, evolutionary changes in heterochiasmy need not always require direct selection on the sex-specific recombination program: chromosome fusion, fission or centromere repositioning may alter the physical extent of male recombination suppression while the underlying sex-specific program remains similar.

This interaction also creates feedback with sex chromosome. By determining the amount of effective intrachromosomal shuffling, pre-existing sex-specific recombination landscapes can influence the extent and rate of sex chromosome differentiation. Yet the contrasting histories of ranids, *Leptobrachium* and *Xenopus* show that this effect is contingent rather than universal. The central challenge is therefore no longer simply to explain why males and females have different recombination rates, but to understand how sex-specific meiotic regulation and evolving chromosome architecture jointly generate, and are subsequently shaped by, sex-specific recombination landscapes.

## Author Contributions

E. U.: conceptualization, data curation, methodology, formal analysis, visualization, writing-original draft, writing-review and editing;

P.V.: supervision, writing-review and editing;

W.-J. M.: conceptualization, methodology, visualization, funding acquisition, resources, supervision, writing-original draft, writing-review and editing

All authors gave final approval for publication and agreed to be held accountable for the work performed therein.

## Acknowledgments

We thank Kim Roelants, Jonathan Goldenberg, Jelter Bekaert for their help in phylogenetic and statistic suggestions and feedback. This work is supported by the European Research Council (ERC) Starting Grant (FrogWY, 101039501), a Research Council grant (OZR4049) from Vrije Universiteit Brussel, and a research grant (G0A3B26N) from the Research Foundation Flanders (Fonds Wetenschappelijk Onderzoek,

FWO), awarded to Wen-Juan Ma. The views and opinions expressed are those of the author(s) only and do not necessarily reflect those of the European Union or the European Research Council Executive Agency. Neither the European Union nor the granting authorities can be held responsible for them.

## Declaration of AI use

AI-assisted technologies were used for language editing, including grammar and spelling checks, and for generating editable shape templates used in figure development. The tools used were ChatGPT (OpenAI) and Claude (Anthropic).

## Conflicts of Interest

The authors declare no conflict of interest.

## Data Accessibility

The comparative dataset compiled from published literature, including recombination characteristics and associated species-level variables for 97 species, together with the phylogenetic data and R scripts used for statistical analyses and figure generation, will be made publicly available in https://github.com/TheWMaLab upon publication. The original data sources are cited in the manuscript and supplementary materials.

# Supplementary Material

# The evolution of extreme spatial heterochiasmy in amphibians: interactions between sex-specific meiotoc programme and chromosome architecture

Ezgi Unal1[†], Paris Veltsos[1], Wen-Juan Ma[1†]

[1.] Research group of Ecology, Evolution and Genetics, Biology Department, Vrije Universiteit Brussel, Brussels, Belgium

[†]Correspondence:

Wen-Juan.Ma@vub.be

Hatice.Ezgi.Unal@vub.be

## Table of Contents

# Supplementary Methods

## S1. Data compilation and dataset construction

Species-level information on recombination patterns, karyotype characteristics, genome size and sex-determining systems was compiled from published cytogenetic and genetic studies. Recombination data were obtained from analyses of male meiotic chromosomes, female lampbrush chromosomes and sex-specific linkage maps, where available. Given the variation in the type and resolution of recombination data among studies, chiasma patterns were standardised according to their distribution along bivalents (terminal or interstitial) and differences between the sexes.

For male meiosis, published cytological images and accompanying descriptions were examined to classify chiasma location. Ring-shaped bivalents at diakinesis or metaphase I were classified as exhibiting terminally localised chiasmata, typically with one chiasma per chromosome arm, whereas non-ring configurations, including rod- and cross-shaped bivalents, were classified as exhibiting interstitial chiasmata. Each bivalent was assigned to one of these two categories, and the percentage of bivalents in each category was calculated relative to the total number of bivalents found per species.

Female recombination information was available for 20 of the 97 species (20.6%). Of these, 17 species had cytological data, including 11 with complementary sex-specific linkage maps, while three were represented by linkage-map data alone. Across these species, female recombination was consistently interstitial and broadly distributed along chromosomes, without evidence of strong terminal localisation. Strong spatial heterochiasmy was inferred in 16 species exhibiting predominantly terminal recombination in males but broadly distributed recombination in females. The remaining four species (*Ascaphus truei*, *Bombina variegata*, *Ambystoma mexicanum* and *Oedipina uniformis*), which lacked pronounced terminal restriction in males, were considered as not exhibiting strong spatial heterochiasmy.

Karyotype characteristics were compiled from published sources, including haploid chromosome number and the proportions of metacentric/submetacentric, acrocentric and telocentric chromosomes. Sex-determination systems were classified according to published evidence as XY, ZW, ZWXY or W0. Genome-size estimates were obtained from NCBI where reported in gigabases, or from the Animal Genome Size Database (Gregory, 2026), with haploid C-values converted from picograms to gigabases using 1 pg DNA = 0.978 Gb.

To maximise taxonomic coverage while maintaining complete information for each statistical model, separate analysis-specific datasets were constructed. Three principal datasets were generated: (i) a male-chiasma dataset comprising 97 species with information on chiasma localisation and karyotype characteristics; (ii) a subset of 54 species with additional genome-size estimates; and (iii) a subset of 40 species with known sex-determination systems. Analyses of sex-determination systems were restricted to XY and ZW species, excluding two observations classified as W0 and ZWXY. Chromosome number and karyotype composition were included in all three datasets.

**S2. Composite phylogenetic tree construction**

Phylogenetic comparative analyses were conducted using a composite time-calibrated phylogeny combining the anuran phylogeny of Portik et al. (2023) with the broader amphibian phylogeny of Jetz and Pyron (2018). The source trees were independently pruned to retain species represented in our compiled dataset. The composite tree preserved the topology, internal branch lengths and timescale of the Portik et al. anuran subtree, while the Jetz and Pyron phylogeny provided relationships within Caudata and Gymnophiona and the deeper topology among amphibian orders (Gymnophiona sister to Batrachia, comprising Anura and Caudata) (Jetz & Pyron, 2018; Portik et al., 2023).

To reconcile the two phylogenetic timescales, we retained the crown age of Anura from Portik et al. (2023) and used the temporal intervals between crown Anura, Batrachia and the amphibian root estimated from Jetz and Pyron (2018). Node ages were calculated by subtracting root-to-node distances from total tree height. The intervals between crown Anura and Batrachia (64.79 Myr), and between crown Anura and the amphibian root (90.65 Myr), were added to the Portik et al. crown-Anura age without rescaling the anuran subtree. This yielded node ages of 179.34 Ma for crown Anura, 244.13 Ma for Batrachia and 269.99 Ma for the amphibian root. The Caudata and Gymnophiona subtrees retained their original internal branch lengths, while connecting branches were adjusted to match these ancestral-node ages.

The resulting phylogeny comprised 97 species (86 Anura, 9 Caudata and 2 Gymnophiona). We verified that the tree was rooted, fully bifurcating and ultrametric within a numerical tolerance of $10^{-6}$, and confirmed the monophyly of all three amphibian orders and Batrachia. Pairwise patristic distances among anurans were preserved to numerical precision following tree assembly.

**S3. Phylogenetic signal and ancestral-state reconstruction of male chiasma patterns**

We quantified phylogenetic signal and reconstructed the evolutionary history of male recombination patterns using the proportion of bivalents exhibiting terminally localised chiasmata as a continuous trait. Analyses were conducted across 97 amphibian species using the corresponding pruned, time-calibrated phylogeny. For each species, the proportion of terminal bivalents was calculated by dividing the number of bivalents classified as terminal by the total number of bivalents. To accommodate proportions equal to zero or one, values were adjusted using the transformation $p^* = [p(N - 1) + 0.5]/N$, where p is the observed proportion and N is the number of species (97). The adjusted proportions were then logit-transformed as $\log[p^*/(1 - p^*)]$.

Phylogenetic signal was quantified using Pagel's $\lambda$ and Blomberg's K. The significance of Pagel's $\lambda$ was evaluated using a likelihood-ratio test against a model with $\lambda=0$, whereas the significance of Blomberg's (K) was assessed using 9,999 randomisations.

Ancestral states were reconstructed from the logit-transformed proportions under a Brownian-motion model using restricted maximum likelihood (REML) with *ace()*, and estimates were compared with those obtained using *fastAnc()*. Ancestral estimates and their 95% confidence intervals were back-transformed to percentages by applying the inverse-logit function and reversing the initial adjustment. Reconstructed values were visualised across the phylogeny, with nine focal ancestral nodes highlighted.

## S4. Phylogenetic comparative analyses of male chiasma patterns

We used phylogenetic generalised linear mixed models (PGLMMs) to examine associations between male chiasma localisation and chromosome morphology, genome size, mean chromosome size, and sex-determination systems. Models were fitted using *pglmm_compare()* with a binomial error distribution, logit link, and a two-column response specifying terminal and interstitial bivalent counts. Phylogenetic relatedness was accounted for using the corresponding pruned phylogeny, with an additional observation-level random effect. Models were fitted using restricted maximum likelihood (REML).

For the complete dataset of 97 species, we examined associations between male chiasma localisation and chromosome morphology, represented by the proportions of telocentric, metacentric/submetacentric and acrocentric/subtelocentric chromosomes. Each chromosome-morphology variable was included in a separate model, with haploid chromosome number included as a covariate.

For the subset of 54 species with available genome-size information, we examined associations between male chiasma localisation, chromosome morphology and genome size. Separate models included either the proportion of telocentric or metacentric/submetacentric chromosomes, together with genome size and haploid chromosome number. We additionally examined mean chromosome size, calculated as haploid genome size (Mb) divided by haploid chromosome number, in separate models including chromosome morphology and haploid chromosome number.

Finally, we investigated associations between male chiasma localisation and sex-determination systems using the subset of 40 species with XY (28 species) or ZW (12 species) systems. Separate models included sex-determination system together with either the proportion of telocentric or metacentric/submetacentric chromosomes, controlling for haploid chromosome number. XY was specified as the reference category.

Continuous predictors were centred and standardised to unit variance before model fitting. Genome size, mean chromosome size and haploid chromosome number were natural-log-transformed before standardisation. Statistical significance of fixed effects was evaluated using Wald z-tests. For graphical presentation, model predictions were calculated from fixed-effect estimates on the logit scale and back-transformed to percentages using the inverse-logit function. Where applicable, 95% confidence intervals were calculated from the fixed-effect covariance matrix. For comparisons between sex-determination systems, model-based means and standard deviations were estimated using 4,000 multivariate-normal draws of the fixed-effect coefficients, with continuous covariates held at their standardised means. Full model results, including coefficient estimates, standard errors and associated significance tests, are provided in the Supplementary Table section.

## S5. Software and reproducibility

All phylogenetic comparative analyses were conducted in *R version 4.6.1 (R Core Team, 2026)*. Phylogenetic tree assembly, manipulation and matching with trait data were performed using *ape (v. 5.8.1)* and *geiger (v. 2.0.12)* (Paradis & Schliep, 2019; Pennell et al., 2014). Phylogenetic signal was assessed using *phytools (v. 2.5.2)*, ancestral states were reconstructed using ape, and phylogenetic generalised linear mixed models were fitted using *phyr (v. 1.1.3)* (Li et al., 2020; Revell, 2024). Data manipulation and visualisation were performed primarily using *dplyr*, *ggplot2* and *ggtree* (Yu et al., 2017). The compiled species-level trait dataset, composite phylogenetic tree and R scripts used to reproduce the analyses and figures are available at https://github.com/TheWMaLab upon publication.

## Supplementary Figures

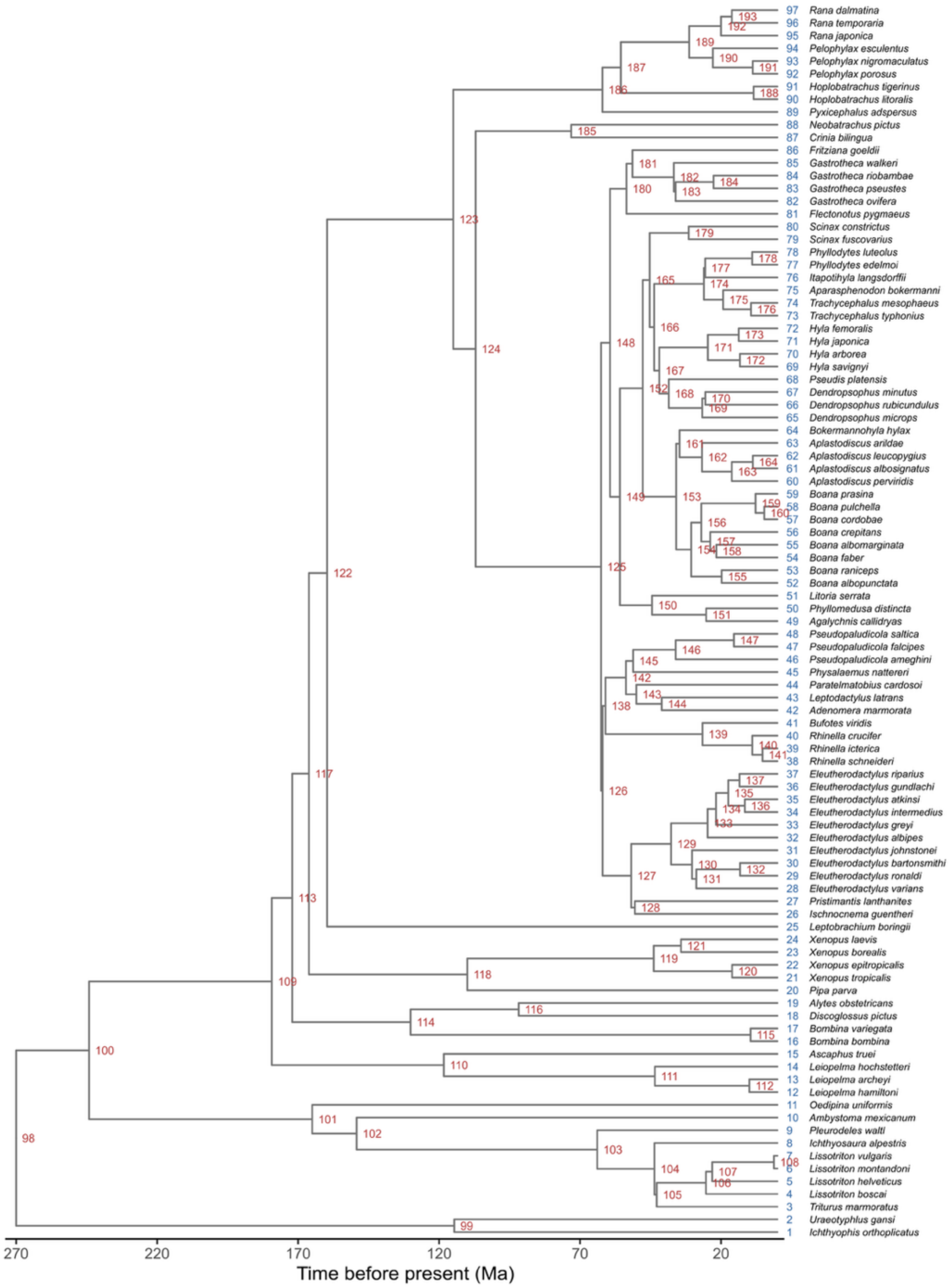

**Supplementary Figure S1. Time-calibrated phylogeny of 97 amphibian species showing node identifiers used in the continuous ancestral-state reconstruction.** Tip node numbers (blue) correspond to individual species, whereas internal node numbers (red) identify ancestral nodes. Branch lengths represent evolutionary time in millions of years before present (Ma). Node identifiers correspond to those used in the accompanying complete table of ancestral-state estimates and associated 95% confidence intervals for the percentage of male bivalents with terminal chiasmata.

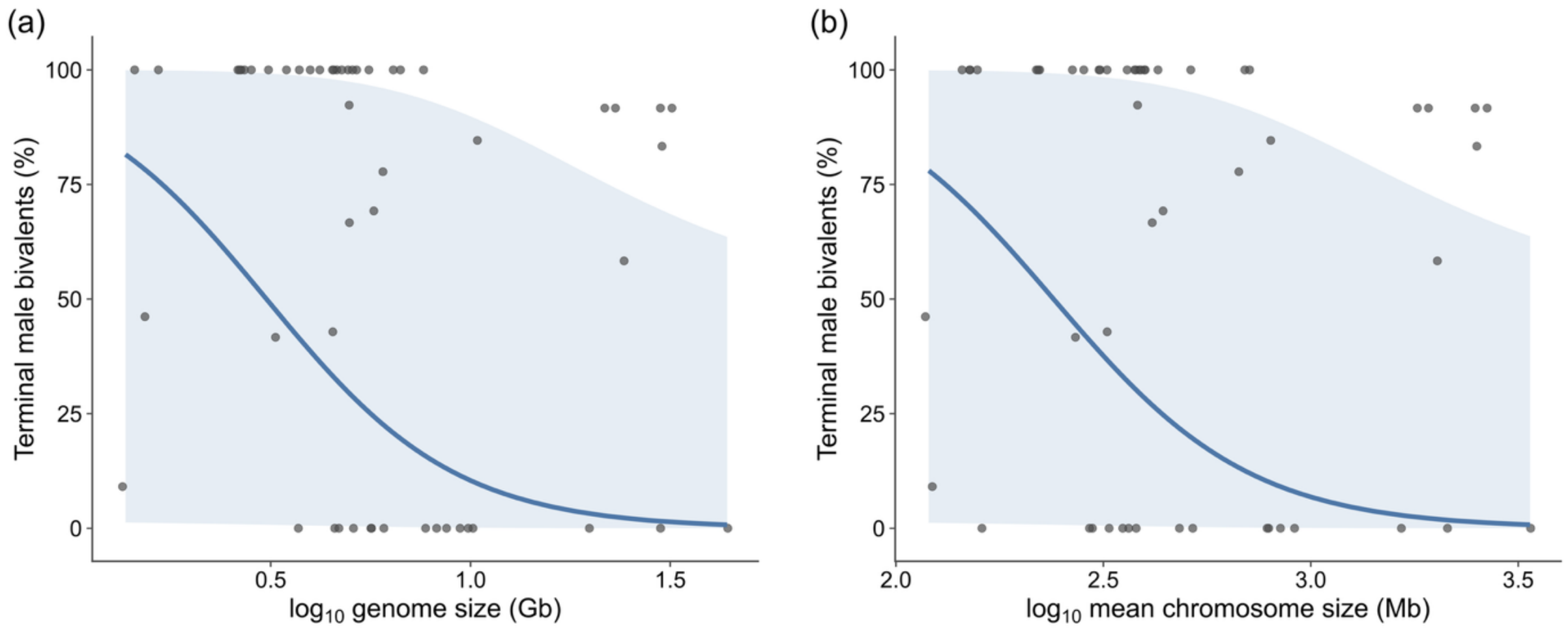

**Supplementary Figure S2. Associations between genome size, mean chromosome size, and male chiasma localization across 54 amphibian species.** (a) Relationship between genome size (Gb) and the percentage of male bivalents with terminal chiasmata. (b) Relationship between mean chromosome size (Mb) and the percentage of male bivalents with terminal chiasmata. Both size variables are displayed on logarithmic scales. Points represent individual species, solid lines indicate predictions from separate phylogenetic binomial generalized linear mixed models, and shaded areas represent 95% confidence intervals around the fitted relationships. Models accounted for telocentric chromosome prevalence, haploid chromosome number, and phylogenetic relatedness. Neither genome size nor mean chromosome size was significantly associated with terminal male chiasma localization ($P = 0.082$ for both models; Supplementary Table S3). Because mean chromosome size was derived from genome size and chromosome number, the two models represent alternative parameterizations rather than independent tests.

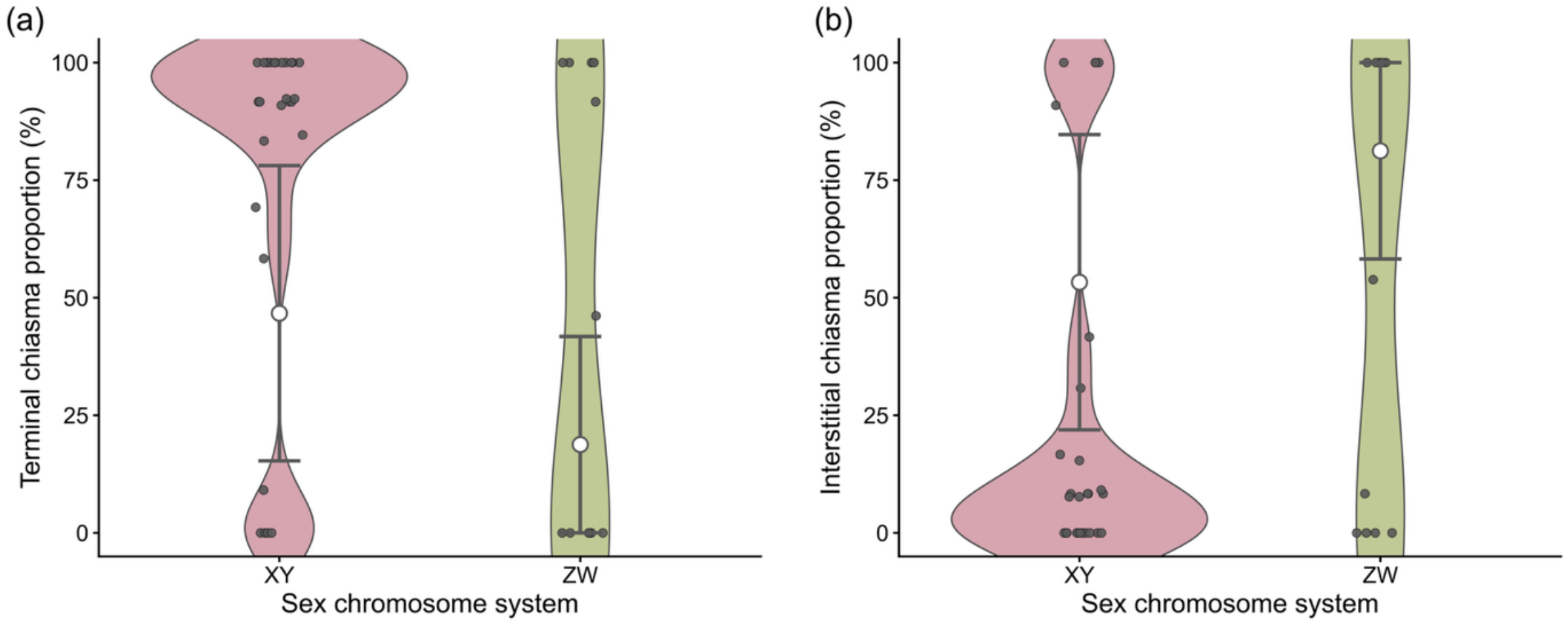

**Supplementary Figure S3. Association between sex chromosome system and male chiasma localization across 40 amphibian species.** (a) Distribution of the percentage of male bivalents with terminal chiasmata and (b) with interstitial chiasmata in species with XY ($n$ = 28) and ZW ($n$ = 12) sex chromosome systems. Violin plots show species-level distributions, and individual points represent species observations. Model-fitted group means and error bars represent predicted percentages ±1 standard deviation, calculated from 4,000 simulated fixed-effect coefficient draws. Predictions were obtained from a phylogenetic binomial generalized linear mixed model accounting for telocentric chromosome prevalence, haploid chromosome number, and phylogenetic relatedness. The association between sex chromosome system and male chiasma localization was not statistically significant ($P$ = 0.072; Supplementary Table S4).

## Supplementary Tables

**Supplementary Table S1. Reconstructed percentages of male bivalents with terminal chiasmata and associated 95% confidence intervals for nine focal ancestral nodes in the phylogeny of 97 amphibian species.**

| Node | Age (Ma) | Estimate terminal chiasma % | Lower 95 % | Upper 95 % |
|---|---|---|---|---|
| 98 | 270 | 4.9 | 0.0 | 99.3 |
| 100 | 244 | 6.6 | 0.0 | 98.3 |
| 109 | 179 | 19.1 | 0.0 | 96.5 |
| 113 | 172 | 25.7 | 0.0 | 96.9 |
| 117 | 166 | 38.3 | 0.3 | 98.4 |
| 122 | 160 | 49.9 | 0.7 | 99.3 |
| 123 | 115 | 85.8 | 8.4 | 100.0 |
| 124 | 107 | 87.7 | 10.6 | 100.0 |
| 125 | 63 | 84.2 | 39.3 | 98.1 |

**Supplementary Table S2a. Phylogenetic models testing the association between chromosome morphology and male recombination topology across 97 amphibian species.** Parameter estimates, standard errors (SE), *z*-statistics, and *P*-values from three separate phylogenetic binomial generalized linear mixed models examining the associations between male chiasma localization and the percentages of (1) telocentric, (2) metacentric/submetacentric, and (3) acrocentric/subtelocentric chromosomes. Each model included chromosome number (n) as a covariate and accounted for phylogenetic relatedness. Chromosome morphology percentages were standardized to a mean of zero and a standard deviation of one. Bivalent number was natural-log-transformed and subsequently standardized. Model coefficients are reported on the log-odds scale.

| **Predictor** | **Estimate** | **SE** | **Z** | **P value** |
|---|---|---|---|---|
| **Model 1: Telocentric % + chromosome number** | | | | |
| Intercept | -2.706 | 2.420 | -1.118 | 0.2636 |
| Telocentric % | -2.130 | 0.392 | -5.437 | 0.0000000541 |
| Chromosome number (n) | -0.112 | 0.458 | -0.244 | 0.8072 |
| **Model 2: Metacentric % + chromosome number** | | | | |
| Intercept | -2.433 | 2.626 | -0.927 | 0.3541 |
| Metacentric /submetacentric % | 2.144 | 0.492 | 4.361 | 0.0000129 |
| Chromosome number (n) | 0.070 | 0.509 | 0.138 | 0.8901 |
| **Model 3: Acrocentric % + chromosome number** | | | | |
| Intercept | -2.397 | 2.840 | -0.844 | 0.3985 |
| Acrocentric/subtelocentric % | 0.716 | 0.275 | 2.601 | 0.0093 |
| Chromosome number (n) | -1.318 | 0.487 | -2.704 | 0.0069 |

**Table S2b. Fixed-effect predictions of the percentage of male bivalents with terminal chiasmata across three chromosome morphology models (97 amphibian species).** Predictions were calculated at -1, 0 and +1 standard deviations (SD) of chromosome morphology prevalence, with standardized chromosome number held at zero. Predicted percentages and 95% confidence intervals (CI) were calculated from the fixed effects, with random effects set to zero. The final column indicates whether the corresponding chromosome morphology percentage falls within the biologically possible range of 0-100%.

| Model | SD | Morphology (%) | Predicted terminal bivalents (%) | 95% CI (%) | Valid range |
|---|---|---|---|---|---|
| **Model 1: Telocentric** | -1 | -16.61 | 36.00 | 0.48-98.50 | *No* |
| | 0 | 12.71 | 6.26 | 0.06-88.48 | Yes |
| | +1 | 42.03 | 0.79 | 0.01-50.31 | Yes |
| **Model 2: Metacentric/submetacentric** | -1 | 48.62 | 1.02 | 0.01-66.48 | Yes |
| | 0 | 78.99 | 8.07 | 0.05-93.78 | Yes |
| | +1 | 109.36 | 42.83 | 0.41-99.28 | *No* |
| **Model 3: Acrocentric/subtelocentric** | -1 | -7.16 | 4.25 | 0.02-92.45 | *No* |
| | 0 | 8.30 | 8.34 | 0.03-95.96 | Yes |
| | +1 | 23.76 | 15.70 | 0.07-97.98 | Yes |

**Limitations:** Because chromosome-morphology percentages are bounded between 0% and 100%, three predictor values corresponding to ±1 SD fall outside the biologically possible range. Three standardized predictor values correspond to biologically impossible chromosome percentages (below 0% or above 100%). Furthermore, the 95% confidence intervals are very wide, indicating substantial uncertainty in the absolute predicted percentages of terminal male bivalents.

**Supplementary Table S3. Phylogenetic models testing the associations of genome size and mean chromosome size with male chiasma localization across 54 amphibian species.** Parameter estimates, standard errors (SE), *z*-statistics, and *P*-values from four separate phylogenetic binomial generalized linear mixed models examining the associations between male chiasma localization and genome size or mean chromosome size. Each model included either telocentric or metacentric/submetacentric chromosome prevalence, together with chromosome number (*n*) as a covariate, and accounted for phylogenetic relatedness. Chromosome morphology percentages were standardized, while genome size, mean chromosome size, and bivalent number were natural-log-transformed before standardization. Model coefficients are reported on the log-odds scale.

| **Predictor** | **Estimate** | **SE** | **Z** | **P value** |
|---|---|---|---|---|
| **Model 4: Telocentric % + genome size + chromosome number** | | | | |
| Intercept | -1.244 | 2.246 | -0.554 | 0.5798 |
| Telocentric % | -3.172 | 0.637 | -4.982 | 0.000000631 |
| Genome size | -1.504 | 0.864 | -1.741 | 0.0817 |
| Chromosome number (n) | 0.183 | 0.636 | 0.288 | 0.7731 |
| **Model 5: Telocentric % + mean chromosome size + chromosome number** | | | | |
| Intercept | -1.244 | 2.246 | -0.554 | 0.5798 |
| Telocentric % | -3.172 | 0.637 | -4.982 | 0.000000631 |
| Mean chromosome size | -1.556 | 0.894 | -1.741 | 0.0817 |
| Chromosome number (n) | -0.156 | 0.644 | -0.242 | 0.8091 |
| **Model 6: Metacentric/submetacentric % + genome size + chromosome number** | | | | |
| Intercept | -1.557 | 2.899 | -0.537 | 0.5911 |
| Metacentric/submetacentric % | 2.531 | 0.947 | 2.672 | 0.0075 |
| Genome size | -1.374 | 1.059 | -1.298 | 0.1943 |
| Chromosome number (n) | 0.027 | 0.853 | 0.031 | 0.9750 |
| **Model 7: Metacentric/submetacentric % + mean chromosome size + chromosome number** | | | | |
| Intercept | -1.557 | 2.899 | -0.537 | 0.5911 |
| Metacentric/submetacentric % | 2.531 | 0.947 | 2.672 | 0.0075 |
| Mean chromosome size | -1.422 | 1.095 | -1.298 | 0.1943 |
| Chromosome number (n) | -0.283 | 0.859 | -0.330 | 0.7418 |

**Supplementary Table S4. Phylogenetic models testing the association between sex chromosome system and male chiasma localization across 40 amphibian species.** Parameter estimates, standard errors (SE), *z*-statistics, and *P*-values from two separate phylogenetic binomial generalized linear mixed models examining the association between sex chromosome system and male chiasma localization. Both models compared ZW with XY sex chromosome systems (reference category: XY), while including either telocentric or metacentric/submetacentric chromosome prevalence and bivalent number (*n*) as covariates and accounting for phylogenetic relatedness. Chromosome-morphology percentages were standardized, while chromosome number was natural-log-transformed before standardization. Model coefficients are reported on the log-odds scale.

| Predictor | Estimate | SE | Z | P value |
|---|---|---|---|---|
| **Model 8: Telocentric % + sex chromosome system + chromosome number** | | | | |
| Intercept | -0.200 | 2.003 | -0.100 | 0.9206 |
| Telocentric % | -2.040 | 0.554 | -3.682 | 0.0002317 |
| Sex chromosome system ZW vs XY | -2.166 | 1.204 | -1.799 | 0.0720 |
| Chromosome number (n) | 0.572 | 0.728 | 0.786 | 0.4319 |
| **Model 9: Metacentric/submetacentric % + sex chromosome system + chromosome number** | | | | |
| Intercept | -0.444 | 2.063 | -0.215 | 0.8295 |
| Metacentric/submetacentric % | 1.833 | 0.557 | 3.293 | 0.0009916 |
| Sex chromosome system (ZW vs XY) | -1.888 | 1.193 | -1.583 | 0.1135 |
| Chromosome number (n) | 0.619 | 0.733 | 0.845 | 0.3983 |

## Supplementary Dataset

**Supplementary Dataset S1. Complete species-level dataset for 97 amphibian species.**

The complete dataset used in the comparative analyses, including male/female chiasma localization and recombination information, chromosome morphology, and associated species traits, is provided separately as a CSV file rather than reproduced as a table in this document.

**Data file:** Supplementary_Dataset_S1.csv

**Supplementary Dataset S2. Complete continuous ancestral-state reconstruction results.**

Reconstructed percentages of male bivalents with terminal chiasmata and associated 95% confidence intervals for all internal nodes of the 97-species amphibian phylogeny. Ancestral states were estimated under a Brownian-motion model using restricted maximum likelihood (REML). Node identifiers correspond to the phylogenetic tree in Supplementary Figure S1.

**Data file:** Supplementary_Dataset_S2.csv

## Supplementary References